\documentclass[12pt,oneside]{amsart} 

\usepackage[T1]{fontenc}%{polish}%[babel]
\usepackage{graphicx}
\usepackage{amssymb, amsmath}
\usepackage{amsfonts}
\usepackage{amssymb}
\usepackage{longtable}
\usepackage{longtable}
\usepackage{pdfpages}

\newtheorem{theorem}{Theorem}

\newtheorem{exercise}[theorem]{Exercise}

\numberwithin{equation}{section}
\numberwithin{theorem}{section}
\numberwithin{example}{section}
\numberwithin{figure}{section}

\begin{document}

\centerline{\bf  How Wave - Wavelet Trading  Wins and ``Beats'' the Market}
\bigskip
\centerline{\bf Lanh Tran}
\bigskip
\centerline{Department of Statistics, Indiana University, Bloomington, U.S.A.}
\bigskip
%  \make title
\centerline{Abstract}
\bigskip
{\it  The purpose of this paper is to showcase trading strategies that give solutions to 
three difficult and intriguing problems in business finance, economics and statistics. The
paper discusses trading strategies for both commodities and stocks but the  main focus
is on  stock market trading at the New York Stock Exchange. 
\vskip .1in
\centerline {  Problem 1: Buy Low and Sell High} 
\vskip .1in
The buy low and sell high problem can be summarized like this:  suppose  the price of a commodity or stock 
fluctuates indefinitely, is there
any explicit strategy for a trader to ``ride the price waves''  by buying low and selling high to eventually win  
even if price does not increase? 
Buy low and sell high is the basic tenet of 
most forms of trading. However, people who engage in this kind of trading usually lose 
since it is not clear what ``high'' and  ``low''  mean and how much to sell when price is
high and how much to buy when price is low. Part 1 of the paper shows how these 
problems can be dealt with by  employing my 
webpage ``Portfolio Checker''  on my website AgateTrading.com.  This strategy can be 
used as a guide for the trading of any stock and a trader using it 
always wins in the long run.  The only condition assumed is that the standard deviation  of the stock 
price traded stays bounded away from zero.
\vskip .1in
\centerline {Problem 2:  ``Beat''  the Market}
\vskip .1in
 The market here mainly refers to  the S\&P 500 index SPY which is an important benchmark of U.S. stock performance.  
The share price of SPY has increased close to  700 percent from January 1993 to December 2016.  The following problem
is of great interest: is there any trading system that can ``beat''  the market consistently? Part 2 of this paper addresses this
question. While the trading system  in Part 1 always wins eventually, it does not ``beat''  the market consistently
since its return rises more slowly than the S\&P 500 benchmark.  In Part 2, the trading 
system presented in Part 1 is transformed into a strategy that always outperforms the market eventually. Using this
modified strategy requires a  computer program on my webpage Wavelet Trading at my website AgateTrading.com.  The 
strategy also works well as 
a guide for trading other stocks with similar traits as the SPY, for example, the 
Russell 2000 Index IWM or SPDR Dow Jones Industrial Index ETF DIA.  
%An interesting 
%consequence is that my findings in Part 2 contradict  the 
%Efficient Market and Random Walk hypotheses, which are considered to be cornerstones of 
%modern finance theory.  Both hypotheses assert that 
%it is impossible for a trader to outperform the market consistently.
\vskip .1in
\centerline{Problem 3: Can a Trader Outperform a geometric Brownian Motion?}
\vskip .1in
The geometric Brownian motion (GBM) is employed in the Black-Scholes model 
and is a very popular model for stock market price.  
GBM is also known as exponential Brownian motion. GBM is always positive 
and tends to infinity with probability 1 if the drift parameter is positive.  In addition, GBM is a Markov process, which 
means that all information on past historical prices is incorporated in the current price.  The general belief is that
it is impossible to ``beat''  a GBM since technical analysis of historical prices is useless in predicting future prices. 
The last part of the paper shows that the answer to Problem 3 is actually a  ``YES'', which is quite surprising.
\vskip .1 in
%The  three problems, in particular Problems 2 and 3, % are  considered to be unsolvable due
%to the unpredictability of stock prices. 
The trading strategies presented are based mainly on information obtained  
from the movements of waves and wavelets  created by large and small fluctuations of market prices. 
They do not involve any forecasting or prediction of future prices.   
Behavioral economics also plays a  role in the decision making process of the Wavelet Trading program. 
My website AgateTrading.com is available to the public.  
}

\bigskip
-----------------------------------------------------------------------------------

Key words and Phrases.  Wavelets, efficient market, geometric Brownian motion.

\vfill\eject

\section{Introduction }\label{sec:Intro} 

Many economists, financial analysts, researchers, traders, and people
acquainted with modern business finance 
theories consider that  trading commodities and stocks by any system of buying and selling to
make a profit is an impossible 
task, as there is plenty of empirical evidence 
and statistics against the existence
of such a strategy;  traders
who buy and sell frequently usually end up losing money. In addition, 
proponents of the random walk and efficient market hypotheses, considered to be 
cornerstones of modern finance theory and economics, 
 assert that stocks take an unpredictable path and that it is 
 impossible to outperform 
 the overall market consistently because stocks always 
 trade at their fair value on stock exchanges. 
 Observing that most people 
lose money in stock market trading brings many to conclude
that trading is purely a game of luck.  This paper will
show that trading commodities and stocks is, in reality, more a game of skill than of luck.  

There is a large body
 of literature on the random walk and efficient market hypotheses.  
 See Basu (1977), Chan, Gup and Pan (2003), Cootner (1964),
 Fama (1965a, 1965b, 1970), Guerrien and Gun (2011), Lo and MacKinlay (1999),  Lo (2005), Malkiel (1973), 
 Pearson (1905),  Samuelson (1965), Schwert (2001), Sharpe (1964, 1975), Timmermann and Granger (2004), 
 and the references therein for an 
 account of this information. The random walk theory became well known
 since the year 1973 when Malkiel (1973) published his book 
 ``A Random Walk Down Wall Street.''    The efficient market hypothesis evolved from the work
 of Fama  (1965a, 1965b).  Both the random walk and
 the efficient market hypotheses have a strong influence on economics, modern market finances,
 and stock market trading.  There are daily articles in newspapers on trading and related topics. The
 list of references listed here is a tiny fraction of the actual amount available.
  
The market is implicitly assumed to be the Standard and Poor's 500 
SPY index, unless indicated otherwise.   The SPY  index represents a  collection of 500 stocks 
of a diverse group of U.S. companies and is a popular benchmark to measure 
U.S.  stock performance.  The SPY data set (Histo SPY Data)  displayed on the website plays an important role in the paper. It 
contains the dates and corresponding adjusted closing prices of  SPY from January 29th, 1993  to December 30th, 2016.
The adjusted  closing prices represent a more accurate reflection of  SPY's performance.  They take into account corporate 
 actions such as stock splits and dividends.
The data was downloaded on January 1st, 2017 from the Yahoo Finance website online. 
These prices may look different if they are downloaded on another day.  Many examples given
in the paper use the data from the SPY data set.   A date of the year will often be displayed
 in the same style of the SPY data set. For example, January 29th, 1993 is written as 1/29/93.  
 
 The tables posted on the website  are integral parts of the paper but they are too long to be included here. 
 Each table will be explained as we go along.
  Interactive Brokers (IB)  is used as the  brokerage in many  examples. The commission charged by IB is \$1.00  if the 
 number of shares per trade is 200 or less; it is 
\$0.005 per share if the trade involves more shares.   All 
money referred to is in U.S. dollars.   
The paper is organized as follows:
\vskip .1in
{\bf Part 1.} The buy low and sell high problem is discussed in this part which contains Sections 2 and 3. 
Section 2 presents the Portfolio Checker program (denoted by PC) on 
my website and gives instructions as to how to employ it as a guide for buying lows and selling highs.  A  lot of effort on my part has 
been devoted to make the  computer program 
user friendly.  Being able to operate the computer program is necessary 
to understand what comes later.   Section 3 discusses the PC strategy in more detail.
\vskip .1in
{\bf Part 2.}  This part investigates the ``Beat the Market'' problem. It contains Sections 4-6. 
Section 4 presents the Wavelet Trading program (denoted by WT) and instructions for a user.  
To evaluate the performance of WT, I use the SPY data set displayed on the Website as a test case. 
Table 1 on the website shows  the trades by WT on each trading day  from 1/29/93 to 12/30/16. 
WT is programmed to always start the trading by buying 1,500 shares on the first trading day. A share of
SPY cost \$28.000838 on this day so WT has to pay \$42,001.26 for the cost of the shares and another
\$7.50 in commission. Note that Table 1 contains  some other vital statistics: commissions, cost, cumulative cost,  and
the profit made by the trader. There are many
days with no trades.  Table 2  is the same as Table 1 with the days of no trades deleted. Table 1 and Table 2 
show that WT made a profit of \$396,639.09 after her latest trade on 12/30/16.  

Did WT outperform a buy-and-hold investor (denoted by BH) who bought 1,500 shares on 1/29/93 and left it 
alone in the brokerage until 12/30/16?  Sections 5 and 6 give an affirmative ``YES''  answer to this 
question.   Assume the following: WT deposited into the brokerage an initial amount 
of \$42,001.26 and \$7.50 in commission on 1/29/93. WT
uses margins to buy shares when necessary according to the rules set up by Interactive Brokers (IB).  Currently,  IB allows
a regular trader to use a 2:1 leverage using margin. 

Under this scenario, the profit  of \$396,639.09 mentioned above is not WT's actual return since it does not take into account 
the interests paid to (or earned from) the brokerage when she owed the brokerage money (or had surplus cash) in 
her account.  Table 3 in Section 5 shows the computation of the returns of WT and BH from the starting date on 1/29/93 to the
last day of trading on 12/30/16.  

Section 6 compares the performance of WT and BH.  Table 3 shows that the returns of WT and BH are, 
respectively,   \$393,169.57 and \$293,286.24.  By trading, WT has earned a return  larger than
BH's return by \$99,883.33 or 34.06 percent.  Did WT earn a higher return by assuming additional risk?  Computing the
Sharpe ratios  of WT and BH and comparing them is a way to answer this question. The yearly rate of return
of WT has a standard deviation of  \$17.57 as compared with \$17.24 for BH and 
the Sharpe Ratios of WT  and BH are found to be 0.44 and 0.38. 
Therefore, WT has obtained much better risk adjusted returns than the market.
The last part of this section shows that WT always ``beats''  the market eventually for any arbitrary starting trading date.  Tables
4 and 5 are examples to persuade the reader that this is indeed the case. 
Section 7 compares PC versus WT and discusses some  important points that were 
not mentioned in previous sections.

In Section 8, a geometric Brownian motion (GBM) is fitted to Histo Spy Data.  
Brownian motion can take on negative values but stock prices cannot.  Thus, 
using Brownian motion to model
stock prices is disputable. However, GBM only takes
positive values and is a Markov process. The future values of GBM given the present value are
independent of past values. GBM is often used to model stock prices in a non-arbitrage world.
Using the fitted GBM,  simulated data is generated to show that  WT  always ``beats'' BH eventually.  
It is quite surprising that a trader can ``beat'' a  market having a perfect fair price by 
utilizing market fluctuations.  At the end of the paper,  a heuristic proof that the market is not efficient will
be given.

%Exercise 7.2 has a very
%interesting consequence: it says that if an economy is fully efficient, 
%then market price will increase in the long run. 

There is a quick way to see
that WT  ``beats'' the market. From Table 1, WT makes \$396,639.09 in profit while BH makes 
$1500\times (\$223.529999-\$28.000838)-\$7.50=\$293,296.2$.   Note that \$7.50 is the commission
paid to buy 1,500 shares on the first day of trading. WT's return  is only a few thousand dollars less if
margin rates of interest are taken into account (see Table 3).

Throughout the paper, I will implicitly assume that the standard deviation 
of market prices stays bounded away from zero, that is, it stays above
some positive number.  
When I use the words ``tends to'', I  always refer to a limit as time tends to infinity.

\vskip .1in
\vfill\eject
  \centerline{\bf Part 1: Buy Low  and Sell High}

 \section{The Portfolio Checker Program}\label{sec:PortfolioChecker} 
 This section explains how to operate the portfolio checker.  The user starts by 
entering  the ``Magic Number''  on the bottom left side of the screen right below the
``Generate Table''  button.  The Magic Number is currently set to be 8141946.
\vskip .1 in
 {\bf Remark 2.1.}{ \it  
 It is important to enter the Magic Number correctly
 on the screen. 
 } \qed
\vskip .05 in
She then needs to come up with three numbers before she 
makes a trade: the current market price of 
a share of  stock being considered for selling or buying, the 
number of shares to buy (a positive integer) or sell (a negative integer) 
and the commission for the trade.
She then asks PC if such a trade is feasible.
This is done by entering the three numbers on the screen with a comma or a space
as a separator and then activating the computer program by clicking on the 
button with the words ``Generate Table''.   On the screen,  don't separate digits
in a number by using commas. For example: write 1250 instead of 1,250.
PC analyzes these numbers and answers the question 
``Is my Portfolio OK?'' with a simple  ``YES'' or ``NO''.  The user needs to
make sure that she gets  a  ``YES''  each time she makes a trade.

The following rules are set by PC:

\vskip .05in

 Rule1.  The first trade must be a  buy of at least 20 shares. In addition, the cost
of the first trade must be at least \$1,000.00. 

Rule 2.  Don't sell if the current price of a share is less than the price at its last trade.

Rule 3.  Don't buy if the current price of a share is higher than the price at its last trade.

 Rule 4.  Don't buy or sell an amount less than 50 times the
commission.

\vskip .05in 

{\bf  Remark 2.2.}  {\it If you buy a stock of \$2.50 a share then
to satisfy Rule 1, you need to start with at least 400 shares, while if you buy a stock of \$250.00 a 
share, then you need to start with at least 20 shares, which cost \$5,000.00.  The number of shares 
you buy in your first transaction depends on the total amount of money 
you have allocated for that particular stock. Some funds should be available for additional purchases 
if necessary.
Rules 2 and 3 are essential elements of the buy low and sell high scheme.  
Rule 4 is to ensure that you don't buy or sell an amount that is too small to justify 
the commission that you pay.  } \qed
%\end{enumerate}

\vskip .05in

PC  does not recommend a number of shares to buy or sell. This gives more flexibility
to the trader in deciding as to how much to buy or sell. 
The computer program is designed to assure the user that she will eventually make a profit
if she gets a ``YES''  answer to the portfolio question 
before  each trade that she makes. The 
trader controls her account balance. If her account balance gets too
low, she should buy more often than sell to raise it up.

A ``NO'' answer occurs due to various  reasons.  The most common ones are:

\vskip .05in 
(1)   The price of the stock traded has not increased enough or decreased enough from the price at
its  last trade to warrant,  respectively,  a ``sell'' 
 or a ``buy''.  A regular trader with 300 shares of SPY and  a bank of about \$100,000.00 should 
 wait for the price to change 
 about 2 percent or more from the price at its last trade before she attempts to trade.

(2) The number of shares entered is  too high or too low.
\vskip .1in 

{\bf Example 2.1.} {\it  Below is an example of  5 trades of the stock 
SPY, made respectively by a trader, on January 2nd, 
6th, 8th, 14th and February 3rd in the year 2015.   Before each trade, the
trader used the Portfolio Checker to make sure that she got ``YES'' for an answer.
The trader's 5 transactions are summarized in the table below:}
\vskip .1 in
\tabcolsep4.0pt
\noindent{\sffamily\tiny
\begin{tabular}{rcrccrccrr}
\textbf{ \,} & \textbf{} & \textbf{} & \textbf{} & \textbf{Cum} & 
\textbf{\,\,\,} & \textbf{Cum} &\textbf{Cum} & \textbf{} & \textbf{} \\
\textbf{Dates \,} & \textbf{Price} & \textbf{Shares} & \textbf{Com} & \textbf{Com} & 
\textbf{Cost\,\,\,} & \textbf{Cost} &\textbf{Shares} & \textbf{MV} & \textbf{Profit} \\
\hline
1/2/15	&	197.05	&	300	&	1.50	         &	1.50	&	59115.00	&	59115.00	&	300	&	59115.00	&	-1.50	\\
1/6/15	&	191.66	&	50	&	 1.00 	&	2.50	&	9583.00	&	68698.00	&	350	&	67081.00	&	-1619.50	\\
1/8/15	&	197.50	&	-36	&	 1.00 	&	3.50	&	-7110.00	&	61588.00	&	314	&	62015.00	&	423.50	\\
1/14/15	&	192.66	&	36	&	 1.00 	&	4.50	&	6935.76	&	68523.76	&	350	&	67431.00	&	-1097.26	\\
2/3/15	&	196.48	&	-28	&	 1.00 	&	5.50	&	-5501.44	&	63022.32	&	322	&	63266.56	&	238.74	\\
\end{tabular}}
\vskip .1in
\qed
\vskip .1in
The first four columns list respectively the trading dates, share prices at which trades are 
made, number of shares traded, and the corresponding 
commissions. A positive number for the shares means a ``buy''  and a negative one means a ``sell''.  

The fifth column (CumCom) lists
cumulative commissions, which are total commissions paid up to the trading day. For example, on 1/8/15, after
she sold 36 shares at the price of 197.50 a share, she has paid a commission 
of  $\$1.50+\$1.00+\$1.00$, which equals \$3.50.

The sixth column (Cost) lists the cost of each trade which equals the number of shares 
traded multiplied by the price at which it is traded. You have
to pay money if it is a positive number and you get back cash if it is a negative number. A 
positive number for cost indicates a buy and a negative number indicates a sell. 

The seventh column (CumCost) lists the cumulative cost paid by the trader. For example, 
after the fourth transaction is made on 1/14/15, the trader's 
cumulative cost is obtained by adding  \$59,115.00,   \$9,583.00,  $-$\$7,110.00 and \$6,935.76, which equals \$68,523.76. 

The eighth column (CumShare) lists the total number of shares held by the trader after each trade. For 
example, after the fifth trade was carried out
on 2/3/15, the trader has a total of shares equal to 322, which is the sum of all numbers in Column 3.

The ninth column lists the market value of the trader on each trading day. The market value (MV) is computed by
multiplying the cumulative number of shares of the trader by the corresponding price.  For example, on 1/14/15, the 
market price of the trader is $\$192.66\times 350$, which equals \$67,431.00.

Finally, the tenth column lists the profit after each trade. The profit made by the trader 
after each trade is given by:
$$
          \hbox{Profit}=\hbox{MV}-\hbox{CumCost}-\hbox{CumCom.}
$$
The profit after the last trade is  \$63,266.56$-$\$63,022.32$-\$$5.50=\$238.74.
Note the interesting fact that the trader has made a profit of \$238.74 while the market price
drops from \$197.05 a share to \$196.48 a share. This example is simple, but it
indicates that the trader can win by riding the ``price waves''. 

\vskip .05 in
The trader was able to utilize fluctuations of the price of the stock to make profits. Let us see what happens if we
arbitrarily replace 191.66 by 189.66,  197.50 by 200.50 and 192.66 by 187.66. The table above would change to:
\vskip .1 in
\tabcolsep4.0pt
\noindent{\sffamily\tiny
\begin{tabular}{rcrccrccrr}
\textbf{ \,} & \textbf{} & \textbf{} & \textbf{} & \textbf{Cum} & 
\textbf{\,\,\,} & \textbf{Cum} &\textbf{Cum} & \textbf{} & \textbf{} \\
\textbf{Dates \,} & \textbf{Price} & \textbf{Shares} & \textbf{Com} & \textbf{Com} & 
\textbf{Cost\,\,\,} & \textbf{Cost} &\textbf{Shares} & \textbf{MV} & \textbf{Profit} \\
\hline
1/2/15	&	197.05	&	300	&	1.50	&	1.50	&	59115.00	&	59115.00	&	300	&	59115.00	&	-1.50	\\
1/6/15	&	189.66	&	50	&	1.00	&	2.50	&	9483.00	&	68598.00	&	350	&	66381.00	&	-2219.50	\\
1/8/15	&	200.50	&	-36	&	1.00	&	3.50	&	-7218.00	&	61380.00	&	314	&	62957.00	&	1573.50	\\
1/14/15	&	187.66	&	36	&	1.00	&	4.50	&	6755.76	&	68135.76	&	350	&	65681.00	&	-2459.26	\\
2/3/15	&	196.48	&	-28	&	1.00	&	5.50	&	-5501.44	&	62634.32	&	322	&	63266.56	&	626.74	\\
\end{tabular}}
\vskip .1in	
\noindent She makes a profit of \$626.74  which is much more than \$238.74 since the price 
of the stock fluctuates more.

It is a good idea for you to  copy and paste the data into a text or Excel file and keep it 
as your record.  If you mistype and then activate the program, the whole data that
you have entered may disappear. In case this happens, you still have a 
record of what you have typed, you can copy and paste your
data back onto the computer screen, correct the mistype,  and then reactivate the program. The 
website does not hold any data for you after you logout.
\vskip .1 in
{\bf Exercise 2.1.}  {\it As an exercise,  you may try to enter the prices, 
shares, commissions in the table
above on the screen as follows: 
197.05, 300, 1.50, 
189.66, 50, 1.00,
 200.50, -36, 1.00, 
 187.66, 36, 1.00, 
 196.48,  -28, 1.00.
 The numbers need to be separated by
 a comma or a space. The
answer  to the portfolio question is a ``YES'' after you activate the program.  
}\qed
\vskip .1in
{\bf Remark 2.3.}   {\it  Following Example 2.1, what should your next move be? 
There are 322 shares left. How about selling ten percent of the
shares (32 shares) if your stock price goes up 2 percent or buying 32
additional shares if your stock price decreases 2 percent.  The stock price
becomes $.98\times \$196.48=\$192.55$ if it decreases 2 percent and 
$1.02\times \$196.48=\$200.41$ if it increases 2 percent.
You find out both of the trades get  a ``YES'' answer by asking PC. Then 
you can make an advance order with your brokerage to perform 
whichever transaction that comes first. }\qed
\vskip .1in
There are many ways to enter the data on the ``DATA TO CALCULATE'' area.  You may find it
more to your liking to  enter the data above as follows:
\vskip .1 truein
197.05, \hskip.022in  300, 1.50 

189.66,  \hskip.097in 50, 1.00 

200.50, $-$36,  1.00 

 187.66, \hskip.14in  36, 1.00 

196.48,  $-$28, 1.00  
 \vskip .05 in
 { \bf Exercise 2.2.} {\it   
Assume that Scottrade is your brokerage and that your bank is \$100,000.00.  
Note  that Scottrade charges a commission fee of 7 dollars for each transaction no matter
how many shares are involved.
Using the computer program and the five prices of the stock SPY above, but  starting with a buy 
of 350 shares on 1/2/2105, find 4 additional 
feasible trades on, respectively, the  6th, 8th, 14th and February 3rd  of 2015.  To get a feasible
trade, you have to get a ``YES'' answer to the portfolio question before each trade is made. Do you
make a profit after you finish with the last trade?
}\qed
 \vskip .05 in
Example 2.1 shows that it is possible for you to make a profit  even if
the share price of the traded stock decreases if you trade properly.  You 
can also lose if the share  price of the traded stock
increases if trading is not done right.  An example is given below.
 \vskip .05 in
{\bf Example 2.2.} {\it  Consider the following list of three transactions 
made by a trader using Interactive Brokers as the brokerage.}
\vskip .1 in
\tabcolsep4.0pt
\noindent{\sffamily\tiny
\begin{tabular}{rcrccrccrr}
\textbf{ \,} & \textbf{} & \textbf{} & \textbf{} & \textbf{Cum} & 
\textbf{\,\,\,} & \textbf{Cum} &\textbf{Cum} & \textbf{} & \textbf{} \\
\textbf{ \,} & \textbf{Price} & \textbf{Shares} & \textbf{Com} & \textbf{Com} & 
\textbf{Cost\,\,\,} & \textbf{Cost} &\textbf{Shares} & \textbf{MV} & \textbf{Profit} \\
\hline
&  75.49	&	1000	&	5.00	&	5.00	&	75490.00	&	75490.00	&	1000	&	75490.00	&	-5.00	\\
&  86.03	&	-100	&	1.00	&	6.00	&	-8603.00	&	66887.00	&	900	&	77427.00	&	10534.00	\\
&  82.97	&	200	&	1.00	&	7.00	&	16594.00	&	83481.00	&	1100	&	91267.00	&	7779.00	\\
\end{tabular}}
\vskip .1in	
{\it After the last transaction is made, she has a total of 1100 shares.  If the share price
drops down to  \$75.60,  the market value of these shares would be 
$1100\times \$75.60=\$83,160.00$,  which is less than the  cumulative cost of \$83,481.00.  She ends up with a loss of 
\$83,481.00-\$83,160.00+\$7.00, which equals, \$328.00, not counting the commission 
she has to pay if she cashes out.  She loses money while the price of a share goes up from
\$75.49 to \$75.60.
} 
\qed
\vskip .1in
Let us ask the computer if these transactions are OK by entering 
\vskip .01 in
 
 75.49, \hskip .05in 1000, 5.00
  
86.03,  $-1$00, 1.00 

82.97,  \hskip .12in 200, 1.00 
 
\noindent on the screen and activate the program. 
The answer is a ``NO''.  The reason is because 
the amount of shares she bought on the last transaction 
is too high.  
\vskip .05in
How much should she buy then?  She should just try a smaller number that she is comfortable with 
and then ask the computer if buying the latter is ok before making such a trade. How about asking the portfolio
checker if buying 60 at \$82.97 is ok? This is done by entering
\vskip .1in
 75.49,\hskip .05in 1000, 5.00
  
86.03,$-1$00, 1.00 

82.97, \hskip .16in 60, 1.00 
\vskip .1in
\noindent on the screen. The answer is ``YES''.  Suppose she buys 60 shares
at  \$82.97 per share, and then 40 shares if the price
 drops back to the starting price of \$75.49, she would  make 
 a profit of \$597.20 as shown in the table below. 
 \vskip .1in
\tabcolsep4.0pt
\noindent{\sffamily\tiny
\begin{tabular}{rcrccrccrr}
\textbf{ \,} & \textbf{} & \textbf{} & \textbf{} & \textbf{Cum} & 
\textbf{\,\,\,} & \textbf{Cum} &\textbf{Cum} & \textbf{} & \textbf{} \\
\textbf{ \,} & \textbf{Price} & \textbf{Shares} & \textbf{Com} & \textbf{Com} & 
\textbf{Cost\,\,\,} & \textbf{Cost} &\textbf{Shares} & \textbf{MV} & \textbf{Profit} \\
\hline
&  75.49	&	1000	&	5.00	&	5.00	&	75490.00	&	75490.00	&	1000	&	75490.00	&	-5.00	\\
&  86.03	&	-100	&	1.00	&	6.00	&	-8603.00	&	66887.00	&	900	&	77427.00	&	10534.00	\\
&  82.97	&	60	&	1.00	&	7.00	&	4978.20	&	71865.20	&	960	&	79651.20	&	7779.00	\\
&  75.49	&	40	&	1.00	&	8.00	&	3019.60	&	74884.80	&	1000	&	75490.00	&	597.20	\\
\end{tabular}}
\vskip .1in	
 She again holds 1,000 shares
 which is the same number of shares she started with. The choice of buying
 60 shares and then 40 is purely arbitrary.  Buying 40 shares at  \$82.97 a share
 and then 60 shares if price falls to \$75.49 would  also get a ``YES'' answer 
 from the Portfolio Checker.
The answer to the question as to how many shares to buy or how many shares
to sell depends on many variables:  the commission you have to pay, cost of a share,
among others.  The best way to learn how to use the website is for you
to play with it.
\vskip .1truein
{\bf Exercise 2.3.}  {\it  In Example 2.2 above, buying 200 shares at \$82.97 a share is not 
an acceptable trade. What is the maximum allowable number of shares that she can buy at \$82.97 to 
get a ``YES'' answer?}
\vskip .1in
{\bf Hint: } {\it Use portfolio checker and get the answer by trial and error. Decrease the number 200 slowly
until you get ``YES'' for an answer.}\qed
\vskip .1in
It is fairly easy for you to get a ``YES''  answer from the portfolio checker after
some practice.  Before you quit using the website, make sure you save the historical
data of your trades  in some file. Click on the ``Select Table''  button on 
the upper right hand corner of the screen and use copy and paste
to save your data. 

You need to upload all the data back
to the screen in case  you want to ask the portfolio checker some time later if 
your ``next'' trade is ok or not.  Whether your ``next'' trade is feasible or not depends
not only on your ``last'' move but on the entire historical data of your trades.
%%Answer is 140
 \section{Can PC ``beat'' the market?}\label{sec:LowsHighs} 
The table below shows 
the trading of SPY from 4/6/98 to 9/8/03 when its share  price moved from \$79.60 on 4/6/98 to \$79.03 on 9/8/03, 
which is a period of over 5 years. During this period, PC makes a profit of  \$14,721.15 while the market
went down slightly. However, market price does not stay flat all the time. It can rise 
and fall dramatically for long periods of time in, respectively, a bull market and bear market.
During these periods
the market (or BH) can ``beat''  PC  badly.   Below are two main problems:
\vskip .1in
1.  {\bf Trader sells too many of her shares when market goes up.} 
{\it Suppose a trader sells 10 percent of her shares whenever her stock price 
goes up 2 percent, she will have few shares left to sell if price goes straight up
by 20 percent.  And if price rises even more thereafter, the market ends up beating
her badly. She would have won more if she had held to her shares without any selling.}
\vskip.1in
2.  {\bf Trader buys  too many of her shares when market goes down.} 
{\it Suppose a trader buys to increase her shares by 10 percent whenever 
her stock price  goes down 2 percent, she will have more than twice
as many shares as she started with if market price goes straight down
20 percent. She can lose big if price keeps falling thereafter.} 
\vskip.1in
As a remedy, use the following rule when you use PC: don't let your cumulative cost vary by a large amount.
This will prevent you from selling or buying too much when price is, respectively, too high or too
 low. 
\vskip .1 in
\tabcolsep4.0pt
\noindent{\sffamily\tiny
\begin{tabular}{rcrccrccrr}
\textbf{ \,} & \textbf{} & \textbf{} & \textbf{} & \textbf{Cum} & 
\textbf{\,\,\,} & \textbf{Cum} &\textbf{Cum} & \textbf{} & \textbf{} \\
\textbf{Dates \,} & \textbf{Price} & \textbf{Shares} & \textbf{Com} & \textbf{Com} & 
\textbf{Cost\,\,\,} & \textbf{Cost} &\textbf{Shares} & \textbf{MV} & \textbf{Profit} \\
\hline
4/6/98	&	79.60	&	1000	&	5.00	&	5.00	&	79600.00     &	  79600.00	&	1000	&	79600.00	&	-5.00	\\
4/27/98	&	77.49	&	100	&	1.00	&	6.00	&	7749.00        &	  87349.00	&	1100	&	85239.00	&	-2116.00	\\
5/20/98	&	80.11	&	-100	&	1.00	&	7.00	&	-8011.00	    &     79338.00	&	1000	&	80110.00	&	765.00	\\
9/2/98	&	71.03	&	200	&	1.00	&	8.00	&	14206.00	    &      93544.00	&	1200	&	85236.00	&	-8316.00	\\
11/4/98	&	80.54	&	-150	&	1.00	&	9.00	&	-12081.00	    &      81463.00	&	1050	&	84567.00	&	3095.00	\\
11/23/98	&	85.65	&	-200	&	1.00	&	10.00	&	-17130.00	   &    64333.00	&	850	&	72802.50	&	8459.50	\\
1/7/99	&	91.29	&	-150	&	1.00	&	11.00	&	-13693.50	    &   50639.50	&	700	&	63903.00	&	13252.50	\\
4/22/99	&	98.26	&	-100	&	1.00	&	12.00	&	-9826.00	&     40813.50	&	600	&	58956.00	&	18130.50	\\
6/2/99	&	93.72	&	150	&	1.00	&	13.00	&	14058.00	  &    54871.50	&	750	&	70290.00	&	15405.50	\\
7/13/99	&	100.88	&	-150	&	1.00	&	14.00	&	-15132.00    &	39739.50	&	600	&	60528.00	&	20774.50	\\
8/4/99	&	94.55	&	150	&	1.00	&	15.00	&	14182.50	&     53922.00	&	750	&	70912.50	&	16975.50	\\
8/25/99	&	100.16	&	-100	&	1.00	&	16.00	&	-10016.00	    &    43906.00	&	650	&	65104.00	&	21182.00	\\
9/22/99	&	94.80	&	120	&	1.00	&	17.00	&	11376.00	&     55282.00	&	770	&	72996.00	&	17697.00	\\
11/5/99	&	100.08	&	-120	&	1.00	&	18.00	&	-12009.60	   &    43272.40	&	650	&	65052.00	&	21761.60	\\
3/17/00	&	107.19	&	-150	&	1.00	&	19.00	&	-16078.50	   &   27193.90	&	500	&	53595.00	&	26382.10	\\
12/1/00	&	96.92	&	200	&	1.00	&	20.00	&	19384.00	&    46577.90	&	700	&	67844.00	&	21246.10	\\
1/30/01	&	101.32	&	-120	&	1.00	&	21.00	&	-12158.40   &	34419.50	&	580	&	58765.60	&	24325.10	\\
3/1/01	&	91.62	&	150	&	1.00	&	22.00	&	13743.00   &  	48162.50	&	730	&	66882.60	&	18698.10	\\
4/2/01	&	84.20	&	150	&	1.00	&	23.00	&	12630.00	   &   60792.50	&	880	&	74096.00	&	13280.50	\\
4/18/01	&	91.42	&	-100	&	1.00	&	24.00	&	-9142.00	&  51650.50	&	780	&	71307.60	&	19633.10	\\
8/30/01	&	83.78	&	100	&	1.00	&	25.00	&	8378.00	&  60028.50	&	880	&	73726.40	&	13672.90	\\
9/25/01	&	75.51	&	125	&	1.00	&	26.00	&	9438.75	&  69467.25	&	1005	&	75887.55	&	6394.30	\\
10/23/01	&	80.83	&	-100	&	1.00	&	27.00	&	-8083.00	&  61384.25	&	905	&	73151.15	&	11739.90	\\
7/9/02	&	71.65	&	150	&	1.00	&	28.00	&	10747.50	&  72131.75	&	1055	&	75590.75	&	3431.00	\\
7/19/02	&	63.48	&	150	&	1.00	&	29.00	&	9522.00	&  81653.75	&	1205	&	76493.40	&	-5189.35	\\
8/14/02	&	69.11	&	-150	&	1.00	&	30.00	&	-10366.50 &	71287.25	&	1055	&	72911.05	&	1593.80	\\
9/23/02	&	62.98	&	150	&	1.00	&	31.00	&	9447.00	&  80734.25	&	1205	&	75890.90	&	-4874.35	\\
11/6/02	&	70.04	&	-100	&	1.00	&	32.00	&	-7004.00	&  73730.25	&	1105	&	77394.20	&	3631.95	\\
1/27/03	&	64.45	&	150	&	1.00	&	33.00	&	9667.50	&  83397.75	&	1255	&	80884.75	&	-2546.00	\\
5/2/03	&	70.80	&	-125	&	1.00	&	34.00	&	-8850.00	&  74547.75	&	1130	&	80004.00	&	5422.25	\\
9/8/03	&	79.03	&	-100	&	1.00	&	35.00	&	-7903.00	&  66644.75	&	1030	&	81400.90	&	14721.15	\\
\end{tabular}}
\vskip .1in
\qed
\vskip .1in
{\bf Exercise 3.1} {\it  Show that you always win eventually by using PC.
Please be reminded that the standard deviation of SPY is assumed to always stay above a positive number.
For your information, the standard deviation of SPY prices on Histo SPY Data is \$47.39.} 
\vskip .1in
{\bf Hint.} {\it In the beginning you can lose if share price drops. However, the assumption on the standard deviation
ensures that share price has to stop dropping after some time.  Thereafter, your gain by buy low and sell high
will eventually surpass this initial loss.}\qed
\vskip .4in
\vfill\eject
 \centerline{\bf Part 2:  Wavelet Trading}
 \vskip -.5in
 \section{The Wavelet Trading Program}\label{sec:computer program}
 
Share prices are written with up to 6 decimals in this section. As an example, suppose that a 
trader started  trading SPY on January 3rd, 2000. She entered the date 
and price of SPY (see Hist SPY Data on the website) on this date on the screen:
\vskip .1in
\tabcolsep40.0pt
\noindent{\sffamily\tiny
\begin{tabular}{rc}
1/29/93	&	\hskip -1in  28.000838  \\
\end{tabular}}
\vskip .1in
She then asks WT what action to take by clicking on the
button with the words ``Generate Table''.
WT analyzes these numbers and gives the following reply:
\vskip .1in
\tabcolsep4.0pt
\noindent{\sffamily\tiny
\begin{tabular}{rcrccrccrr}
\textbf{ \,} & \textbf{} & \textbf{} & \textbf{} & \textbf{Cum} & 
\textbf{\,\,\,} & \textbf{Cum} &\textbf{Cum} & \textbf{} & \textbf{} \\
\textbf{Dates \,} & \textbf{Price} & \textbf{Shares} & \textbf{Com} & \textbf{Com} & 
\textbf{Cost\,\,\,} & \textbf{Cost} &\textbf{Shares} & \textbf{MV} & \textbf{Profit} \\
\hline
1/29/93	&	28.000838	&	1500	&	7.50	&	7.50	&	42001.26	&	42001.26	&	1500	&	42001.26	&	-7.50	\\
\end{tabular}}
\vskip .1 in
\noindent 
Now she moved to the second day of  trading and entered the dates and prices of SPY of both days:
\vskip .1in
\tabcolsep40.0pt
\noindent{\sffamily\tiny
\begin{tabular}{rc}
1/29/93	&      \hskip-1in  28.000838\\
2/1/93	&      \hskip -1in 28.199990\\
\end{tabular}}
\vskip .1in
\noindent 
and then clicked on Generate Table. WT replied with the following table:
\vskip .1in
\tabcolsep3.0pt
\noindent{\sffamily\tiny
\begin{tabular}{rcrccrccrr}
\textbf{ \,} & \textbf{} & \textbf{} & \textbf{} & \textbf{Cum} & 
\textbf{\,\,\,} & \textbf{Cum} &\textbf{Cum} & \textbf{} & \textbf{} \\
\textbf{Dates \,} & \textbf{Price} & \textbf{Shares} & \textbf{Com} & \textbf{Com} & 
\textbf{Cost\,\,\,} & \textbf{Cost} &\textbf{Shares} & \textbf{MV} & \textbf{Profit} \\
\hline
1/29/93	&	28.000838	&	1500	&	7.50	&	7.50	&	42001.26	&	42001.26	&	1500	&	42001.26	&	-7.50	\\
2/1/93	&	28.199990	&	0	&	0.00	&	7.50	&	0.00	&	42001.26	&	1500	&	42299.99	&	291.23	\\
\end{tabular}}
\vskip .1in
\noindent WT's  advice on 2/1/93 is not to buy any shares and WT computes the commission, cumulative 
cost and other statistics for the trader.  She kept 
on repeating this process and her record for the first 15 days is:
\vskip .1in
\tabcolsep2.0pt
\noindent{\sffamily\tiny
\begin{tabular}{rcrccrccrr}
\textbf{ \,} & \textbf{} & \textbf{} & \textbf{} & \textbf{Cum} & 
\textbf{\,\,\,} & \textbf{Cum} &\textbf{Cum} & \textbf{} & \textbf{} \\
\textbf{Dates \,} & \textbf{Price} & \textbf{Shares} & \textbf{Com} & \textbf{Com} & 
\textbf{Cost\,\,\,} & \textbf{Cost} &\textbf{Shares} & \textbf{MV} & \textbf{Profit} \\
\hline
1/29/93	&	28.000838	&	1500	&	7.50	&	7.50	&	42001.26	&	42001.26	&	1500	&	42001.26	&	-7.50	\\
2/1/93	&	28.199990	&	0	&	0.00	&	7.50	&	0.00	&	42001.26	&	1500	&	42299.99	&	291.23	\\
2/2/93	&	28.259704	&	0	&	0.00	&	7.50	&	0.00	&	42001.26	&	1500	&	42389.56	&	380.80	\\
2/3/93	&	28.558465	&	0	&	0.00	&	7.50	&	0.00	&	42001.26	&	1500	&	42837.70	&	828.94	\\
2/4/93	&	28.677956	&	0	&	0.00	&	7.50	&	0.00	&	42001.26	&	1500	&	43016.93	&	1008.18	\\
2/5/93	&	28.658009	&	0	&	0.00	&	7.50	&	0.00	&	42001.26	&	1500	&	42987.01	&	978.26	\\
2/8/93	&	28.658009	&	0	&	0.00	&	7.50	&	0.00	&	42001.26	&	1500	&	42987.01	&	978.26	\\
2/9/93	&	28.458857	&	0	&	0.00	&	7.50	&	0.00	&	42001.26	&	1500	&	42688.29	&	679.53	\\
2/10/93	&	28.498687	&	0	&	0.00	&	7.50	&	0.00	&	42001.26	&	1500	&	42748.03	&	739.27	\\
2/11/93	&	28.638126	&	0	&	0.00	&	7.50	&	0.00	&	42001.26	&	1500	&	42957.19	&	948.43	\\
2/12/93	&	28.419026	&	200	&	1.00	&	8.50	&	5683.81	&	47685.06	&	1700	&	48312.34	&	618.78	\\
2/16/93	&	27.702077	&	300	&	1.50	&	10.00	&	8310.62	&	55995.69	&	2000	&	55404.15	&	-601.53	\\
2/17/93	&	27.682194	&	0	&	0.00	&	10.00	&	0.00	&	55995.69	&	2000	&	55364.39	&	-641.30	\\
2/18/93	&	27.662247	&	0	&	0.00	&	10.00	&	0.00	&	55995.69	&	2000	&	55324.49	&	-681.19	\\
2/19/93	&	27.761855	&	-300	&	1.50	&	11.50	&	-8328.56	&	47667.13	&	1700	&	47195.15	&	-483.48	\\
\end{tabular}}
\vskip .1in
{\bf Exercise 4.1.}  {\it Table 1 on the website shows  the number of shares traded by WT on each trading day  from 1/29/93 to 12/30/16.  
Table 2  is the same as Table 1 with the days of no trades deleted.  Go to the Wavelet Trading page of the website AgateTrading.com
and paste the SPY data set onto the screen and activate the program. Download the files resultTable.csv and resultTableNoZeros.csv 
to your computer and verify that they are identical to Tables 1 and 2.} \qed

\section{How Table 3 is established. }\label{sec: NumericalResults}

Assume that both WT and BH have a brokerage account with Interactive Brokers. 
An owner of a  memorandum account with a minimum balance of \$25,000
can use a 2:1 margin to buy additional shares. A trader with more than
\$100,000 can open a  special memorandum account, which allows a  2.25:1 margin.
The strategies of WT and BH are as follows:  WT bought 1,500 shares of SPY at the price of \$28.000838 on 1/29/93. 
The total cost is \$42,008.76  which covers \$42,001.26 for the cost of the shares bought and \$7.50 in commission.
 WT may use margin to buy additional  shares according to the rules set up by Interactive Brokers. 
BH took the same action as WT on 1/29/93 but BH made no other trades from there on.  Note that  the actions of WT and  BH are identical on the first day of trading.

The performances of WT and BH will be compared in the
next section. Table 3 plays an important role in this comparison. 
It is important to understand how the numbers in the table are computed and what the
variables in each column in the table mean. 
The table contains all WT's trades from 1/29/93 to 12/30/16.  The numbers on
Table 2 are used to make Table 3.  The reader should know by now how
the first ten columns  A-I are. The rest of this section is to
explain how the numbers in other columns are computed. 
\vskip .1in
{\bf IRMB and IRMA.}  IRMB denotes the rate of interest 
that the trader pays the brokerage  when she borrows money and IRMA is the rate of interest that she
collects from the brokerage when she has surplus cash in her account.  The rates of interest earned vary with
 different brokerages but currently Interactive Brokers charges about the lowest interest. If the trader has a lot of 
 cash in the account, then she can move it into a sweep account 
 and possibly collect higher interests than leaving it in the brokerage account. To 
make things simple, I just assume that when the trader 
has cash in the account, she collects interest from her brokerage at the rate  equal to half the 
interest rate she pays when she borrows. The rate of interest 
on money borrowed at Interactive Brokers for 2016 is:
\vskip .1in
\tabcolsep40.0pt
\noindent{\sffamily\tiny
\begin{tabular}{ccrr}
\textbf{} & \textbf{} \\
\textbf{\$25K} & \textbf{\hskip-1in \$200K}&\textbf {\hskip-1in \$1.5M } &\textbf {\hskip-.7in \$2.5M } \\
1.30\%& \hskip-1in	1.23\% 	& \hskip-1in	1.16\%    & \hskip-.7in	1.01\%                          \\
\end{tabular}}
\vskip .1in

Interactive Brokers was established in 1993 and is the largest U.S. electronic brokerage firm 
by number of daily average revenue trades.  
I do not have the exact data on interest rates charged by Interactive Brokers for the years before 2016.  The 
rates of interest listed below for the year 1993-2015 are set 
to be 1.05 times the interest rates set by the Federal Reserve board.  These interest rates will 
be employed to calculate the interests 
the trader has to pay on margin when she borrows 
money from Interactive Brokers.   Changing them by a few percentage points 
does not affect much the conclusion of
my findings. From here on, when I write an interest rate of, say 0.013, I mean 1.3\%.
\vskip .1in
\centerline{
\tabcolsep40.0pt
\noindent{\sffamily\tiny
\begin{tabular}{rcc}
\textbf{} & \textbf{} \\
\textbf{Year} & \textbf{\hskip-1in IRMB}&\textbf {\hskip-1in IRMA }\\
1993	& \hskip-1in	0.060 	& \hskip-1in	0.030\\
1994& \hskip-1in	0.050	& \hskip-1in      0.025 \\
1995	&\hskip-1in	0.060	& \hskip-1in	0.030\\
1996	&\hskip-1in	0.065	& \hskip-1in	0.033\\
1997	&\hskip-1in	0.065	& \hskip-1in	0.033\\
1998	&\hskip-1in	0.065	& \hskip-1in	0.033\\
1999	&\hskip-1in	0.065	& \hskip-1in	0.033\\
2000	&\hskip-1in	0.065	& \hskip-1in	0.033\\
2001	&\hskip-1in	0.065	& \hskip-1in	0.033\\
2002	&\hskip-1in	0.070	& \hskip-1in	0.035\\
2003	&\hskip-1in	0.040	& \hskip-1in	0.020\\
2004	&\hskip-1in	0.035	& \hskip-1in	0.018\\
2005	&\hskip-1in	0.050	& \hskip-1in	0.025\\
2006	&\hskip-1in	0.060	& \hskip-1in	0.030\\
2007	&\hskip-1in	0.060	& \hskip-1in	0.030\\
2008	&\hskip-1in	0.050	& \hskip-1in	0.025\\
2009	&\hskip-1in	0.040	& \hskip-1in	0.020\\
2010	&\hskip-1in	0.020	& \hskip-1in	0.010\\
2011	&\hskip-1in	0.020	& \hskip-1in	0.010\\
2012	&\hskip-1in	0.020	& \hskip-1in	0.010\\
2013	&\hskip-1in	0.020	& \hskip-1in	0.010\\
2014	&\hskip-1in	0.020	& \hskip-1in	0.010\\
2015	&\hskip-1in	0.020	& \hskip-1in	0.010\\
2016	&\hskip-1in	0.013	& \hskip-1in	0.006\\
\end{tabular}}
}
\vskip .1in
{\bf Interest.} Interest (denoted by Int) on Table 3 lists the interest that trader has to pay or earns depending, respectively, 
on whether her ``cash'' is positive or negative. Next are some examples:

On 3/25/93, trader owed an interest of \$14.05 on a debt of 
\$14,243.35 held for 6 days. The rate of interest used to calculate interest is 0.06.  

 On 7/20/99,  trader collects \$38.58 in interest for the amount of \$20,632.61 in cash that she held in her
 account for 21 days. The rate of interest used to calculate interest is now only 0.033. \qed
 
 I will later explain more on how interest is calculated
 when I discuss ``Cash''. 
\vskip .1in
{\bf Margin}.  Buying on margin means borrowing money from Interactive Brokers to buy stocks. The 
collateral for the borrowed funds is the stocks and cash in  the investor's account. Before buying 
on margin, an investor needs to open a margin account with the brokerage. Maintenance margin 
refers to the minimum amount of money that must exist in the account before the broker forces 
the investor to deposit more money.

In the United States, the Federal Reserve Board regulates the amount of margin that an 
investor must pay for a security. An investor is required  
to fund at least 50\% of a security's purchase price with cash. She may borrow the 
remaining 50\% from a broker or a dealer.
 \qed
 
{\bf Cash}.  The amount  of cash in  BH's account is always zero since all of BH's money is in stocks. 
The amount of cash in WT's account  varies and is rather complicated to figure out. If she 
buys on margin and owes the brokerage money, the cash amount  is a negative number. She has 
to pay interest on this cash that she owes. 
The ``Cash'' variable of WT is zero between the day 1/29/93 until 2/12/93 
when she bought on margin 200 shares at the price of  \$28.419026 a share. The cost 
is  $200\times \$28.419026$, which  equals \$5,683.81.  Her cash after 
she bought 200 shares is $-$\$5,684.81 which covers the \$1.00 commission and the cost of the 200 shares. The 
negative sign indicates that she owed money 
after making this trade and she had to pay interest on this amount while she kept it. 

Her next trade was a ``buy''   of 300 shares made on 2/16/93 at the price of \$27.702077 a share.  She has to  pay \$1.50 
in commission and \$8,310.62 for the cost  of the shares.  She now owes 
the interest on the \$5,684.81 that she borrowed on 2/12/93.  This 
period is four days long and the interest is 
$$
                {\$5684.81\times 4\times  0.060\over 365}=\$3.74
$$
After the third transaction, her cash is 
$$
          -5684.81-1.50-8310.62-3.74, 
  $$
  which equals -\$14,000.67.
  
Let us compute her cash after one more transaction.  On 2/19/93, she sold 300 shares at the price of  \$27.761855
a share, which amounts to \$8,328.56. Her cash is equal to 
$$
          -14000.67-1.50+8328.56-6.90=-5680.51, 
  $$
This amount consists of \$14,000.67 in debt (carry over), an additional commission of \$1.50, an amount  \$8,328.56 from 
the sale of 300 shares and \$6.90 in interest  on the debt of \$14,000.67 which she held for 3 days.
\qed
 \begin{exercise}\label{interestcomputing}
 As an exercise, you may try to verify (using a calculator) that the trader's cash is \$35.84 after she made the next trade, which 
 is a sell of 200 shares on 3/2/93 at the price of \$28.638126	a share.
 \end{exercise}\qed

{\bf Buying Power.}  This is the money currently available for WT to buy additional shares.
This amount is set to be at most twice the net liquidation value (NLV) of her account.
Her NLV is her account balance which is the sum of her market value (MV)  and cash.  Hence, her buying power 
is 
$$
2NLV -MV=2(MV+\hbox{Cash})-MV= MV+2\hbox {Cash}. 
$$
\vskip .1in
{\bf The Difference between Dates}.  DD on Table 3  denotes  the difference between two trading dates. They are used in the computation of
interests that WT pays if she borrows money from the brokerage to buy additional stocks. They are also used to compute  the interest that
the trader collects if she has surplus cash in her account. \qed

                \section{Comparing the performances of The Trader versus the Buy-and-Hold Investor. }

\label{sec:TraderversusBuyandHoldInvestor}
Table 3 shows that by 12/30/16,   the returns of WT and BH are, respectively,   \$393,169.57 and \$293,286.24. 
Thus WT's return exceeds BH's return by \$99,883.33 or 34.06 percent.
  
Has WT earned much higher risk-adjusted returns than BH?  This question can be answered by finding  
their Sharpe ratios.  The Sharpe Ratio (denoted by SR) is widely used for risk-adjusted returns 
and can be computed using the formula:
 $$
 \hbox {SR} =(\hbox {Average Rate of Returns}-\hbox{Risk-free Rate})/{\sigma(\hbox{Rate of Returns})},
 $$
where $\sigma$ denotes the standard deviation.  The 
numbers on the table below are derived from Table 3.   They are used to compute Sharpe ratios of WT and BH. 
\vskip .1in
\centerline{
\tabcolsep2.0pt
\noindent{\sffamily\tiny
\begin{tabular}{rcrccrc}
\textbf{ \,} & \textbf{} & \textbf{} & \textbf{} & \textbf{} & 
\textbf{\,\,\,} & \textbf{}  \\
\textbf{Year \,} & \textbf{WTStartBal} & \textbf{WTReturn} & \textbf{WTRateRet} & \textbf{BHStartBal} & 
\textbf{BHReturn\,\,\,} & \textbf{BHRateRet}  \\
\hline
 1993	&	42001.26	&	4579.87	&	10.90	&	42001.26	&	3719.15	&	8.85	\\
1994	&	46581.13	&	-1.56	&	0.00	&	45720.41	&	340.25	&	0.74	\\
1995	&	46579.57	&	19311.42	&	41.46	&	46060.66	&	17463.02	&	37.91	\\
1996	&	65891.00	&	16974.05	&	25.76	&	63523.68	&	14192.55	&	22.34	\\
1997	&	82865.04	&	32492.11	&	39.21	&	77716.23	&	26519.12	&	34.12	\\
1998	&	115357.15	&	33274.78	&	28.85	&	104235.36	&	28617.96	&	27.46	\\
1999	&	148631.93	&	31577.01	&	21.25	&	132853.32	&	25884.68	&	19.48	\\
2000	&	180208.93	&	-16876.58	&	-9.37	&	158738.00	&	-16666.83	&	-10.50	\\
2001	&	163332.35	&	-14027.71	&	-8.59	&	142071.17	&	-13020.08	&	-9.16	\\
2002	&	149304.64	&	-30833.54	&	-20.65	&	129051.09	&	-25709.91	&	-19.92	\\
2003	&	118471.10	&	36104.18	&	30.48	&	103341.18	&	28913.34	&	27.98	\\
2004	&	154575.28	&	9996.73	&	6.47	&	132254.52	&	7410.30	&	5.60	\\
2005	&	164572.01	&	17259.24	&	10.49	&	139664.82	&	11876.21	&	8.50	\\
2006	&	181831.25	&	24145.61	&	13.28	&	151541.03	&	19661.98	&	12.97	\\
2007	&	205976.87	&	18983.66	&	9.22	&	171203.01	&	8605.79	&	5.03	\\
2008	&	224960.53	&	-71904.53	&	-31.96	&	179808.80	&	-61701.09	&	-34.31	\\
2009	&	153055.99	&	34505.11	&	22.54	&	118107.72	&	29213.84	&	24.73	\\
2010	&	187561.10	&	26250.15	&	14.00	&	147321.56	&	22667.79	&	15.39	\\
2011	&	213811.25	&	5443.52	&	2.55	&	169989.35	&	2551.37	&	1.50	\\
2012	&	219254.77	&	34422.63	&	15.70	&	172540.72	&	29499.46	&	17.10	\\
2013	&	253677.39	&	67590.80	&	26.64	&	202040.18	&	56096.71	&	27.77	\\
2014	&	321268.19	&	48898.74	&	15.22	&	258136.88	&	37430.89	&	14.50	\\
2015	&	370166.93	&	11880.98	&	3.21	&	295567.78	&	-376.74	&	-0.13	\\
2016	&	382047.91	&	53130.42	&	13.91	&	295191.04	&	40103.96	&	13.59	\\
\end{tabular}}}
\vskip .1in 
I now explain how the numbers on the table are computed.
\noindent WT's account balance is equal to the sum of her stock market value 
and available cash.  BH's account balance is equal to her stock market value (MV).

The starting account balance of WT  in a year is her  account balance at
the first trading day of that year. During the period 1993-2015, the ending account balance of each year 
is set to be the starting account balance of the following year. For 2016, the ending account balance is set 
equal to her account  balance on the last trading day of the year.  Using Table 3, the ending account balance
as of 12/30/2016 is 
found to be \$447,060.00-\$11,881.67 or  \$435,178.33. The return of a certain year  is obtained by 
subtracting her beginning account balance from her ending account balance. 

The 
starting and ending account balances of BH are defined similarly. Note that from Table 3, the ending account 
balance of BH  for 2016 is   $1500\times \$223.529999$ or \$335,295.00. 
The account balances of WT and BH at the start of
 a year are abbreviated, respectively, by WTStartBal and BHStartBal. The yearly 
 returns of WT and BH are denoted, respectively,  by WTReturn and BHReturn.  WT's
  and BH's rates of return are denoted, respectively, by WTRateRet and BHRateRet.

As an example,  I  calculate  the returns and rates of returns for WT and BH for the year 1998.  
The return of WT for this year is equal to   
 $\$148,631.93-\$115,357.15=\$33,274.78$. Note that \$148,631.93 is the ending
 account balance of WT for 1998, which equals the beginning 
 balance of 1999. Her beginning account balance for 1998 is  \$115,357.15. Thus 
 WT's rate of return (WTRateRet) for 1998 equals
                            $$
                            100\times 33274.78 /115357.15=28.85
                            $$
 BH's rates of returns are simpler to calculate.  BH's return for 1998 is equal to                           
$\$132,853.32 - \$104,235.36=\$28,617.96$, and hence BH's rate of return for this year equals
                         $$
                            100\times 28617.96/104235.36=27.46
                            $$
I use 4\% for the risk-free rate. The Sharpe ratios of the rates of returns for WT and BH  are found, respectively, to be
 0.44 and 0.38.  They show that WT  produces much better risk-adjusted returns than BH. \qed

Next, I show that WT eventually outperforms the market using any day
as her first day of trading. Table 4 calculates the returns of WT and BH using 7/16/07 as the starting date. 
This date is about the worst one for WT to start her trading since the market crashed soon thereafter.  However, 
by 12/30/16, her return is \$196,410.94 
as compared with \$145,112.45 for BH. The return of WT exceeds that of BH by 35.35 percent. Table 5 
calculates the returns of WT and BH using 1/4/10 as the starting date of trading.  This date was chosen rather
arbitrarily. WT's return is \$244,508.53
as compared with \$187,965.94 for BH. WT's return exceeds BH's return by 30.08 percent. \qed

 \section{Discussions and Comparisons Between PC and WT}

\label{sec:DC}

1.  Parts 1 and 2 present two different strategies.  Since a share of SPY currently costs over 200 dollars, using 
WT requires a large initial amount of money.  WT is designed mainly to trade the SPY index or other index stocks
that represent a broad section of companies, for example, the
Russell 2000 Index IWM or SPDR Dow Jones Industrial Index ETF DIA.  Using PC would be  a much 
better strategy for trading volatile stocks that do not exhibit any trends, for example,  the Silver Standard Resources Inc.
stock SSRI, the SPDR Gold Shares GLD, or VelocityShares Daily Inverse XIV. Seasonal stocks often exhibit this type of behavior.
\vskip .1in 
2.  PC requires that a trader waits until the price of the stock traded  gets sufficiently low to 
buy or sufficiently high to sell. The waiting time to make the next trade may be long.  A user needs some
skill and practice to familiarize with the strategy before doing real life trading.  
\vskip .1in 
3.  WT uses the movements of daily prices of SPY to make trading decisions. 
A trader using WT must enter the date and price of SPY of every trading day. She is required to buy, sell 
or hold according to the advice of WT.  No practice or skill is needed to operate WT. 
\vskip .1in 
4. Suppose WT advises a trader to buy 200 shares at some price. The trader 
may not be able to buy 200 shares at exactly the
price recommended since the price of the stock might have  changed by a small amount
by the time she actually buys.  Buying at prices approximately equal to the  prices recommended 
by WT would not affect the trader's overall return much.
\vskip .1in 
5. What if  the user of WT trades more frequently, for example, every 
hour of the day?  Trading here means she enters the date, share price 
and take the action recommended. This will probably increase her return.
 It would be interesting to see how WT performs with continuous data. I do not have a definite
answer to the question at this time.
\vskip .1in
6. It is possible to create a sophisticated trading system which can monitor stock prices 
and trade automatically. This could even lead to a higher return.
\vskip .1in 
7.  WT can be adapted to use for any brokerage. A user can also start with 
an amount of shares different than 1,500. 
Table 6 shows the trades of WT starting with 750 shares and using Scottrade as the brokerage.  Note 
that the number of shares of each of WT's trade has been cut down by a half.  Scottrade charges much higher margin rates
of interest than Interactive Brokers. The numbers on the IRMB and IRMA columns of Table 6 are 
only rough estimates of Scottrade's actual
rates since I do not have historical data on  its rates of interest. Her return exceeds the BH's 
return by 17.15 percent.  Scottrade charges \$7.00 in commission per trade and high interest rates 
so WT's return has been reduced substantially. 
\vskip .1in 
8. A hedge fund manager using a 5:1 margin leverage can increase her return substantially by using WT. Her buying
power is now 
$$
5NLV -MV=5(MV+\hbox {Cash})-MV= 4MV+5\hbox {Cash}.
$$
Table 7 lists her trades from 1/29/93 to 12/30/16. She starts with 1,500 shares 
but the number of shares of each of her trade
equals to  twice  the number of shares recommended by WT. Her return
is \$493,753.61 as compared with \$293,286.24 for BH.   By trading, her return 
exceeds the BH's return by 68.35 percent.   Table 8 shows that she never got a 
margin call even if she started her trading on 9/28/07 before the crash of 2008-2009. 
Note that, with a 5:1 leverage, she never got a margin call since her buying power 
is always positive. 
\vskip .1in
9. A trader  gets a margin call if the buying power of her account becomes negative. The probability 
that WT gets a margin call is set very low.  A stock market crash much worse than 
the crash in 2008-09 must occur for WT to get a margin call.  Note that the 
buying power of WT is always a positive number  in Tables 3-9. 
\vskip .1in
10. A regular trader can buy and sell using a more aggressive strategy than WT
(for example, by following the strategy of the hedge fund manager mentioned on point 8 above)
to increase her return. The probability that she ever gets a margin call is still very small. However, rare
events can happen after a long period of time. For example, if she started trading on 7/17/07 using the 
hedge fund manager's strategy mentioned above,  she would likely get a margin call 
by September 2009 if the brokerage only allowed her  a 2:1 margin leverage. 
%If she gets a margin call, then the 
%following contingency plan can be employed: 
% sell down to 1,500 shares and start all over again.
\vskip .1in
11.  WT is established under the assumption that SPY has an upward trend tending to infinity.  WT generally 
outperforms the market  during periods when market price  is increasing. 
As a tradeoff, it might underperform  the market during periods when market price is decreasing.  
\vskip .1in
12. How long does it take for WT to ``beat''  the market?  A bear market can last for years.  Thus it can take years for
WT to outperform  the market.   In Table 4, WT started trading on 7/17/07 at the beginning of a bear market.  It took  5 years of 
trading for her return to exceed that of BH.
\vskip .1in
13. The important matters affecting WT's trades are:  ``How high are the margin interest rates?'', ``How much 
commissions does the brokerage charge?'', ``How much margin leverage is WT allowed to use?''
\vskip .1in
14. Behavioral economics plays an important  role in WT's decision making process.  WT  often 
buys high  when there is a big drop 
in share price. For example, WT bought 500 shares on 9/9/16 at share 
price \$210.964611 but did not buy any shares
on 7/7/16 at share price  \$207.255322 which is a lot less (see Table 1 on the website).  Note 
that the drop in price on on 9/9/16 was much bigger
than the drop in price on 7/7/16.  
\vskip .1in
Below are some exercises for the reader.  Assume reasonable commissions and rates of margin interest.
\vskip .1in
{\bf Exercise 7.1.} {\it Download the historical data of the Russell index 2000. Use Table 3, show 
that \hbox{WT}  always ``beats''  the Russell index in the long run. }
\qed
\vskip .1in
  {\bf Exercise 7.2.}  {\it Suppose that an efficient market ``in the sense that no trader  can outperform it''  exists. 
   Show that its price tends  to infinity.}
\qed
\vskip .1in
{\bf Hint:}  {\it  It is not hard to see from Exercise 3.1 that the return of PC actually 
tends to infinity. Since no trader can ``beat'' an efficient market in the long run, 
BH's return has to tend to infinity at least a bit faster than PC's return.   Note that BH's return can only tend to infinity
if the market price tends to infinity.}
\qed
\vskip .1in
This exercise has an interesting consequence. It shows that stocks whose prices fluctuate widely without a clear 
upward or downward trend are inefficient. 
\qed
\vskip .1in
 {\bf  Exercise 7.3.}  {\it  Persuade yourself that if the market price tends to infinity, then WT ``beats'' the market. You can do this 
 by taking  long sequences of historical prices of index stocks 
with increasing trends   and use the
Wavelet Trading page together with Table 3 to show that PC eventually earns a higher return than BH. }
\qed
\vskip .1in
%{\bf  Exercise 7.4.}  {\it  From Exercises 7.2 and 7.3, 5can you conclude that an efficient market  as described 
%%%5in Exercise 7.2 does not exist? 
% }\qed
   \section{Can Wavelet Trading ``Beat''  a Market that is a Geometric  Brownian Motion?}
 
The equation for a geometric Brownian motion (GBM)  is given by:
$$
S_t=S_0\exp\Big ((\mu-{\sigma^2\over 2})t+\sigma W_t\Big ),
$$
where $W_t$ is standard Brownian motion.  Here, $S_t$  is the value of GBM at time $t$ and $S_0$ is the initial value.  
 GBM serves as an important example of  a stochastic process satisfying a stochastic differential equation. The 
 parameters  $-\infty<\mu<\infty$ and $\sigma>0$ are constants. The mean and variance of $S_t$ are:
\begin{equation}\label{meanofSt}
\mathbb {E} (S_{t})=S_{0}\exp(\mu t), 
\end{equation}
\begin{equation}\label{varianceofSt}
{Var} (S_{t})=(S_{0})^{2}\exp( {2\mu t})\left(\exp ({\sigma ^{2}t})-1\right),
\end{equation}
\vfill\eject
The question raised at the beginning of this section is answered in 3 steps:
\vskip .1in
Step 1.  Fit a GBM to Histo SPY prices.
\vskip .1in
Let $X(t) = \log S_t- \log S_{t-1}$. Then 
$$
X(t)=\mu-(\sigma^2/2)+\sigma(W_t-W_{t-1}).
$$
Then $X(t)$ is normally distributed with mean $\mu-(\sigma^2/2)$ and standard deviation $\sigma$.
A total of 6,025 values of $X_t$ is obtained from Histo SPY Data.  Using these values, estimates of  the
drift parameter $\mu-(\sigma^2/2)$ and standard deviation $\sigma$ are, respectively, 0.000345 and 
0.011772.  These estimates are, respectively, the sample mean and sample standard deviation of
6,025 observations of $X_t$'s. The parameter $\mu$ is estimated by 0.000414.

%$$
%S_t=223.529999\exp\Big (0.000345t+0.016  W_t\Big ).
%$$
Since  the drift parameter $\mu-(\sigma^2/2)$ is positive,  $S_t$ 
tends to infinity with probability 1.  This fits well with 
 SPY which has been rising fairly steadily with time.  The reader is referred to Oksendal, B. K. (2002)
and Ross (2014) for more information on GBM.  \qed
 \vskip .1in
Step  2.  Generate Simulated Data Using the Fitted GBM.
 \vskip .1in
 I now show you  how  to employ the page GBM Sim Data (an Excel file) displayed on the website to generate
simulated prices.  The day 12/30/2016 is set to be $t_0$ and the initial price $S_0$ is set equal to the price of SPY on
this day. The equation for the geometric Brownian motion obtained is
\begin{equation}\label{fittedGBM}
        S_t=223.529999\exp\Big (0.000345t+0.011772 W_t\Big ).
        \end{equation}
Column 1 on GBM Sim Data lists 10,000  dates with 12/30/2016 as the first date. 
Column 2 lists the 10,000 time points from 0 to 9,999.
The first number on  column 3 is the value of standard Brownian motion 
at time 0, which is zero. The second number on column 3
is obtained by generating a standard normal random variable using the 
function NORMINV(RAND(),0,1) in Excel.  The third number on column 3
is obtained by adding the second number to another 
randomly generated standard normal random variable, and so on until the last number, which 
is the sum of 9,999 independent standard normal
random variables. The 3rd column contains 10,000 observations  of simulated Brownian motion.  Column 4 contains
the values of simulated SPY prices computed using (\ref{fittedGBM}).

A simple computation shows that the expectation and variance of $S_{9999}$ 
are, respectively, 
$$
223.529999\exp ({ 0.000414\times 9999})=14032,
$$
and
$$
(223.529999)^2\exp (8.279172)\times (0.385661)=75940958.
$$
\qed

%$$
%Var(GBM_{t})=(223.529999)^{2}\exp ({2\times 0.000473\times 9999})\left(\exp ({(0.016^2)\times 9999})-1\right),
%$$$
\vskip .1in
Step 3.  Evaluate the Performance of WT Using Simulated Data.
\vskip .1in
After you have a list of simulated data, paste your data with dates and prices onto the Wavelet Trading page 
of the website and click on the Generate Table button.  Wavelet Trading does the necessary computations and
produces the output files resultTable.csv and resultTableNoZeros.csv. Download both the files to
your computer.   File resultTable.csv lists all of WT's trades from day 0 to day 9999.  This file includes many days with no trading and
is for your inspection only. File resultTableNoZeros.csv contains only days with trades.

Table 9 on the website displays the returns of WT and BH  using a set of simulated data 
that I obtained. I assume WT is a regular trader  having a 2:1 margin leverage.   
I use the current margin of interest 0.013 as estimates for future years though interest rate will probably 
get higher in the future.  I also use the current commissions set by Interactive Brokers as
estimates of future commissions. After 10,000 days of trading, WT's return is  $\$21,181,912.67$ as 
compared with $\$15,867,769.14$ for BH. Thus, WT's return 
exceeds BH's return by 33.49 percent.  
 WT still outperforms BH by a substantial amount if  the rate of margin 
 interest 0.013 is changed to 0.06, for example. 
  \qed

How many shares should a trader start with?  The default number is set to be 1,500. If 
a trader  can easily borrow money to deposit in the brokerage 
in case she gets a margin call, then she can start with  a number smaller than 1,500, say, 1,000 shares.   
Decreasing the initial number of shares bought results in an increase in the probability of 
a margin call. 
\qed
\vskip .1in
Using Sim Spy Data, the wavelet page, and Table 9, you can give a proof by  simulation to show 
that WT always eventually outperforms BH. This is done by repeatedly 
generating samples of simulated observations and comparing the performances 
of WT and BH for each generated sample. More detail is provided in the exercise 
below.
\vskip .1in
 {\bf Exercise 8.1.} {\it Show that  WT generally outperforms BH after 10,000 days by going through the following steps:}
\vskip .1in
 (i)  {\it Download the Excel file ``GBM Sim Data''  posted  on my website AgateTrading.com to your computer and 
generate 10,000 simulated values of SPY. } 
\vskip .1in
(ii) {\it  Paste your data containing only dates and prices of  GBM on the 
Wavelet Trading page and click on the Generate Table button. Download 
the output files resultTable.csv and resultTableNoZeros.csv to your computer. Before you paste 
your data on the Wavelet Trading page, make sure that 
the trading dates are written with two numbers for the month, two 
numbers for the day and four numbers for the year.  Do the 
trades on  the two output files match?}
\vskip .1in
(iii)   {\it Find the returns of WT and BH for your set of simulated data assuming that
 the margin interest rate is 0.013 and the margin leverage is 2:1. Does WT outperform BH? 
The page ``Computation of Returns''  on the website is an Excel file to help you compute
the returns of WT and BH.} 
 \vskip .1 in
 (iv)  {\it Repeat (i), (ii), (iii) over and over to show that WT outperforms BH.}
\qed
\vskip .1in
\centerline{ A HEURISTIC PROOF THAT SPY IS NOT EFFICIENT}
\vskip .1in
Actually, the proof follows from Exercises 7.2 and 7.3. Recall that an efficient market is as defined in Exercise 7.2.  I will show that either PC or WT
will beat the market consistently. Therefore the market cannot be efficient. 

By the assumption made in the introduction, the standard deviation of the market price
stays above a constant, say 10, for example.   If the market price does not tend to infinity, then there must be some price that it will visit infinitely often. In this case,
PC will ``beat''  the market again and again,  making the market inefficient.

Thus market price  has to tend to infinity to stay efficient. However, in this scenario WT will ``beat''  it. In this section I fit SPY with a GBM.  
The result that WT's return increases at a faster rate than BH's return would not change if SPY is fitted with another stochastic process. \qed

In the proof, I assume that the commissions stay reasonable and a trader should be able to get a margin loan at leverage 2:1.
The Black-Scholes model assumes that it is possible to borrow any amount of money at the risk free rate,
and trades do not incur any fees. If this is the case, WT can increase her return considerably by employing a more aggressive strategy,
for example, the strategy of the hedge fund manager mentioned in Point 8 of Section 7.

\vfill\eject

\vskip .1in

Lanh T. Tran

Department of Statistics, Indiana University

Bloomington, Indiana 47408

Tel: 812-325-2067

email: LanhTran14 at gmail dot com


\begin{thebibliography}{99}


\bibitem{} Basu, Sanjoy (1977).  Investment Performance of Common Stocks in Relation 
to Their Price-Earnings Ratios: A test of the Efficient Markets Hypothesis.  {\it Journal of Finance}  {\bf 32}  663-682

\bibitem{} Chan, Kam C.; Gup, Benton E.; Pan, Ming-Shiun (2003). International Stock Market Efficiency and Integration: 
A Study of Eighteen Nations.  {\it Journal of Business Finance \& Accounting} {\bf  24}  803-813.


\bibitem{}  Cootner, P.  ed. The Random Character of Stock 
Market Prices (Cambridge, MA: M.I.T. Press, 1964).

\bibitem{} Fama, E.  F. (1965a). The behavior of 
stock market prices. {\it Journal of Business} {\bf 38}  34-105.

\bibitem{}  Fama, E.  F. (1965b). Random walks in stock prices. 
{\it Financial Analysts Journal}  {\bf 21}  55-59.

\bibitem{} Fama, Eugene F. (1970). Efficient capital markets: A review 
of theory and empirical work. {\it Journal of Finance} {\bf 25} 383-417.

\bibitem{} Guerrien B. and  Gun,  O. (2011).  Efficient Markets Hypothesis: what are we talking about?, .
{\it Real-World Economics Review} {\bf  56}  19-30, 


\bibitem{} Lo, A.  W. and MacKinlay A. C. (1999). A 
Non-Random Walk Down Wall Street, (Princeton: 
Princeton University Press).

\bibitem{}  Lo, A.  W. (2005). Reconciling efficient markets with behavioral finance: the 
adaptive markets hypothesis.  {\it Journal of Investment Consulting}  {\bf 7}  21-44.


\bibitem{} Malkiel, B.  G.  (1973). A Random Walk 
Down Wall Street (New York: W. W. Norton \& Co., 1st edition.)

\bibitem{}   Oksendal, B. K. (2002), Stochastic Differential Equations: An Introduction with Applications, Springer, p. 326.


\bibitem{} Pearson, K. (1905). The Problem of the Random Walk.  Nature, 72-294.

\bibitem{}   Ross, S.  M. (2014).  Variations on Brownian Motion. Introduction to Probability Models (11th ed.). Amsterdam: Elsevier. pp. 612Ð14. 


\bibitem{} Samuelson, P. (1965). Proof that properly anticipated prices fluctuate randomly.  {\it Industrial Management Review} {\bf  6}  41-9.

\bibitem{} Schwert, G. W. (2001). Anomalies and 
Market Efficiency, in G. Constantinides, et. al., 
Handbook of the Economics of Finance, North Holland.

\bibitem{} Sharpe, William F. (1964). Capital asset prices: A theory of market equilibrium under conditions of risk. 
{\it Journal of Finance} {\bf 19}  425-442.

\bibitem{}  Sharpe, William F. (1975) Adjusting for Risk in Portfolio Performance 
Measurement.   {\it Journal of Portfolio Management} {\bf 29-34}.


\bibitem{}  Timmermann,  A.  and  Granger, C.W.J. (2004). Efficient market hypothesis and forecasting. 
{\it International Journal of Forecasting}  {\bf 20}  15-27




\end{thebibliography}
\end{document}